\documentclass{article_saj}
\usepackage{graphicx,saj,multicol,subeqnarray}
\usepackage{float}
\usepackage{xcolor}
\usepackage{widetext}
\usepackage{url}
\usepackage{bm}
\usepackage{tikz} 
\usepackage{pifont} 
\usepackage{amsfonts}
\usepackage{amssymb}
\usepackage{amsmath,upgreek}
\usepackage{titlesec}
\usepackage{minted}
\titlelabel{\thetitle.\quad}
\definecolor{xlinkcolor}{cmyk}{1,0.6,0,0}
\usepackage[bookmarks=false,         
     pdfnewwindow=true,      
     colorlinks=true,    
     linkcolor=xlinkcolor,     
     citecolor=xlinkcolor,     
     filecolor=xlinkcolor,  
     urlcolor=xlinkcolor,      
final=true]{hyperref}

\renewcommand{\thefootnote}{\fnsymbol{footnote}}

\begin{document}
\parindent=.5cm
\baselineskip=3.8truemm
\columnsep=.5truecm
\newenvironment{lefteqnarray}{\arraycolsep=0pt\begin{eqnarray}}
{\end{eqnarray}\protect\aftergroup\ignorespaces}
\newenvironment{lefteqnarray*}{\arraycolsep=0pt\begin{eqnarray*}}
{\end{eqnarray*}\protect\aftergroup\ignorespaces}
\newenvironment{leftsubeqnarray}{\arraycolsep=0pt\begin{subeqnarray}}
{\end{subeqnarray}\protect\aftergroup\ignorespaces}

\begin{strip}

{\ }

\vskip-1cm

Current Version: October 6, 2025

{\ }
\title{Concentrated N-dimensional AMM with Polar Coordinates in Rust}
\vskip3mm
\authors{Vasily Tolstikov, Marcus Wentz, Joseph Schiarizzi, Derek Ding}
\vskip3mm
\vskip3mm
\Email{vasily@orbwap.org, marcus@orbswap.org, joseph@orbswap.org, derek@orbswap.org }
\vskip3mm

\summary{
We expand on the recent development of n-dimensional automated market makers for stablecoins by showing a way to build concentrated liquidity positions with ticks in polar coordinates in Rust, including the featured ability to skew said concentrated liquidity, while also highlighting the risk of stacking too many stablecoin pools and how to hedge said risk.
}

\end{strip}

\tenrm

\section{INTRODUCTION}

\indent

\footnotetext[0]{Acknowledgments: special thanks to Dan Robinson for liquidity fingerprint discussions and Ciamac Moallemi for explaining stablecoin pool liquidity skew phenomenon.}

There exist two schools of thought for liquidity concentration in Automated Market Makers (AMM). One approach involves the specification of a custom curve with parameters controlling the shape of how liquidity is distributed \cite{ref_url1} such as Curve \cite{ref_url2}, CavalRe \cite{ref_url3}, and Eulerswap \cite{ref_url4}. The second approach consists in giving the Liquidity Provider (LP) the freedom to select a discrete price range where liquidity can be concentrated such as Uniswap v3 \cite{ref_url5} and Orbital AMM \cite{ref_url6}. By combining both we herein focus on an n-dimensional AMM for stablecoins where we:

\par\hang\textindent{(1)} Explain the Orbswap invariant. 
\par\hang\textindent{(2)} Explain limitations of symmetric ticks and propose a mechanism to shift liquidity.
\par\hang\textindent{(3)} Demonstrate swap function in polar coordinates.
\par\hang\textindent{(4)} Concentrate ticks in polar coordinates.
\par\hang\textindent{(5)} Highlight the fragility of multi-dimensional stablecoin pools and show a path for mitigation.
\par\hang\textindent{(6)} Introduce multi-modal liquidity fingerprints.

\renewcommand{\thefootnote}{\arabic{footnote}}

\parindent=.5cm

\indent

\section{Orbswap Invariant}

Uniswap v2 introduced a continuous uniform liquidity distribution \cite{ref_url1} allowing for the trading of a token in the price range of zero to infinity. Yet prices are not infinite nor bound by the zero barrier, to address part of this problem Uniswap v3 was introduced to allow for the sharp concentration of liquidity based on an LP's preferred viewpoint on the range of price movement between zero and infinity \cite{ref_url5}. To achieve a softer concentration of liquidity we can adjust the Uniswap v3 AMM invariant equation below by allowing the curve to fold in on itself. A side effect of it is the ability to access negative prices \cite{ref_url7} which can be disabled for our stablecoin use case.
\begin{equation}
\sum_{i=1}^{n} (x_i - l)^2 = l^2
\label{eq:1}
\end{equation}
where $l=2+\sqrt{2}$ is an offset parameter to pin liquidity to the axes. The liquidity fingerprint of this Concentrated Circular Market Maker (CCMM) is derived in Appendix B and can be extended to create a Concentrated Super-Elliptical Market Maker (CSEMM) \cite{ref_url8} to further widen/narrow and skew the liquidity allocation with the invariant becoming:
\begin{equation}
\sum_{i=1}^{n} \left| \frac{x_i}{\alpha_i} - 1 \right|^{\eta(\alpha_i)} = 1
\label{eq:2}
\end{equation}
where $\eta(x)=\frac{ln(2)}{ln(\frac{x}{x-1})}$. Parameters $\alpha_i$ allow for the skew of liquidity in the tails. By setting $\alpha_i$ to the same number, decreasing their values towards two one can approach the liquidity concentration of a Constant Sum Market Maker (CSMM) $x+y=1$, by decreasing further to $\alpha_i=-1$ one can recover the Constant Product Market Maker (CPMM) $x*y=1$, by widening $\alpha_i$ towards infinity one increases the tails of the distribution until it converges to the Logarithmic Market Scoring Rule (LMSR) invariant \cite{ref_url1}. At $\alpha_i=2+\sqrt2$ we get the CCMM from equation (1) which is the version that we have launched on Arbitrum and expanded the stablecoin pool dimensions $n=6$ on our github \cite{ref_url9}.

\subsection{Liquidity Limitations with Circles}
\begin{figure*}[ht!]
\centerline{\includegraphics[width=0.73\textwidth, keepaspectratio]{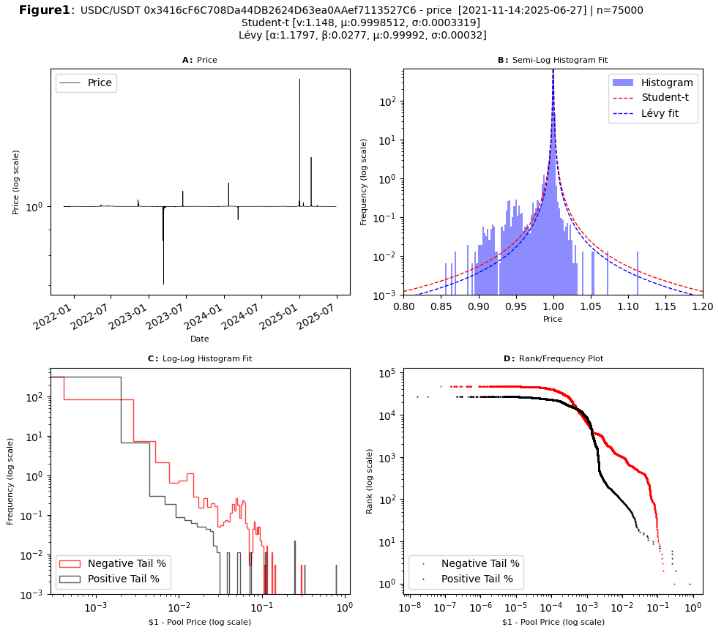}}
\caption{ 1A: Price behavior of USDC/USDT on Uniswap v3 since inception. 1B: Notice the asymmetric skew of the price movement, indicating that providing liquidity to the left of the price makes more sense rather than the right. A personal discussion with Ciamac Moallemi led to the conclusion that the reason for this skew is the redemption fee barrier imposed on one of the stablecoin issuers, which created an incentive to swap the stablecoin on Uniswap into another token without the redemption fee. 1C and 1D in log-log hint at fat-tailed behavior of stablecoin pools. Data retrieved from \href{https://www.dune.com}{Dune Analytics} from the period since the pool's inception on Uniswap v3 until 2025/06/27.}
\label{fig1}
\end{figure*}

The Orbital AMM \cite{ref_url5} provides symmetric ticks, yet liquidity can be demonstrated to be asymmetric in stable pools [Figure~\ref{fig1}], making it capital inefficient to not adjust liquidity. We can construct shifts in the liquidity fingerprint while concentrating liquidity at a specific price peak $c$ by modifying the circular invariant as follows:
\begin{equation}
\left(x-l\right)^{\beta}+\left(\frac{y}{c}-l\right)^{\beta}=l^{\beta}
\end{equation}
where parameter $\beta=2$ for an elliptical case, which is shown to have computational benefits \cite{ref_url10}, and is available in desmos \cite{ref_url11}. It can be extended to $1<\beta<2$ for further concentrating liquidity around a particular price. Each $\beta$ value corresponds to a specific center-curve $C(x)$ for the superellipse
\begin{equation}
C(x)=\frac{1}{1-(1-(\frac{x-1}{x})^\beta)^{\frac{1}{\beta}}}
\end{equation}
along which a scale invariant trading function can be constructed and along which its LP payoff, given via Fenchel conjugacy \cite{ref_url12},  and its liquidity fingerprint, explained in appendix 3.1, shifts [Figure~\ref{fig2}]. 

\subsection{Polar Swap Function}

Unlike traditional AMM swap functions that operate in the Cartesian plane by moving along a curve, a side effect of circular invariants is that it allows one to rotate using polar coordinates \cite{ref_url13} from a focal point with distance $l$ [Figure~\ref{fig3}]. This rotation involves a particular step where the quantity of one stablecoin $\Delta_{x}$ is exchanged for the quantity of a second stablecoin $\Delta_{y}$ while passing through a trigonometric conversion outlined in Rust in Appendix 3.3. 

\subsection{Ticks in Polar Coordinates}
The additional step of trigonometric conversion for Orbswap can be used to construct ticks \cite{ref_url5} in the polar coordinate system based on a particular granularity level [Figure~\ref{fig3}, 4B]. For the sake of simplicity, if our pool initiation consists of n-tokens all priced initially at \$1.00, then the polar tick $\phi$ sits at 45 degrees and has the ability to range from 0 to 90 degrees with the price-to-angle conversion formula being:
\begin{equation}
\phi=\frac{90}{Price+1}
\end{equation}
 As the swap function passes through a particular tick, it can adjust parameter $L$ to change the price impact. Such a discretization approach, when unwrapped [Figure~\ref{fig5}D], happens to be an excellent fit for stable pairs given that ticks are more granular at the body of a price distribution while also being able to capture fat tails, allowing for more accurate capture of price diffusion [Figure~\ref{fig4}].

\subsection{Risk of N-dimensional Pools}
In the event of a permanent stablecoin depeg we witness significant divergence loss for the LP and an accumulation of the unwanted depegged stablecoin. For assets with n-dimensional pools, the first $n>2$ stablecoin example being Curve's Tricrypto pool (USDC,USDT,DAI), one can historically recall the event of Silicon Valley Bank in 2023 \cite{ref_url14} where a bankrun forced USDC to \$$0.85$, resulting in a serious run on the pool and a flight to USDT. The LP of this $n=3$ pool, in the event of a total depeg would have suffered significant divergence loss had the peg not returned. Extending this concept to $n$ stablecoin pools towards infinity poses a significant problem, considering it is building in fragility \cite{ref_url15} of assets that are inherently fat tailed [Figure~\ref{fig5}] and in some cases may even anti-correlate as we've seen with USDC and USDT during the bank run.

To counter the fragility of adding more stablecoins to a pool, we suggest not exceeding such high dimensions, but if one wishes to expand to $n>3$ tokens and include stablecoins with a higher chance of depegging, then LPs could consider hedging their position by constructing a synthetic binary payoff directly with ticks.

The way to replicate the equivalent of a prediction market in the event of a depeg in an N-dimensional pool is by constructing one LP position out of range by one degree and a second LP position that one shorts right below it [Figure~\ref{fig6}] via a vertical spread \cite{ref_url16}. This gives us a good approximation for depeg insurance by mimicking a prediction market for a depeg directly inside of Orbswap by utilizing the existing unused LP positions.

\indent

\section{Appendix}

\subsection{Orbswap Liquidity Fingerprint}
\noindent The liquidity fingerprint can be viewed as the second derivative with respect to the square root of the price \cite{ref_url1}. 
In Cartesian tick space $t$ we get the following distribution of liquidity for the CCMM:
\begin{equation}
    L_{CCMM}(t)=\frac{2le^{\frac{3t}{2}}}{\left(1+e^{2t}\right)^{\frac{3}{2}}}
\end{equation}
which can also be seen in [Figure~\ref{fig2}D] for the $\$1$ price peak. The elliptical case has a similar liquidity fingerprint:
\begin{equation}
    L_{CEMM}\left(t\right)=\frac{2c^{2}le^{\frac{3}{2}t}}{\left(c^{2}+e^{2t}\right)^{\frac{3}{2}}}
\end{equation}

The superelliptical case has only a closed-form solution when $\alpha=\beta$:

\begin{equation}
    L_{CSEMM}\left(t\right)=\frac{2\alpha}{\eta-1}\cdot\frac{e^{\frac{\eta+1}{2\left(\eta-1\right)}\left(t-\ln\left(\frac{s_{y}}{s_{x}}\right)\right)}}{\left(1+e^{\frac{\eta}{\eta-1}\left(t-\ln\left(\frac{s_{y}}{s_{x}}\right)\right)}\right)^{\frac{\eta+1}{\eta}}}
\end{equation}

The value function of an LP payoff is given by the Legendre transform with the Greeks for Delta=$\Delta$, Gamma=$\Gamma$, and Theta=$\Theta$ being available in desmos \cite{ref_url17}. 

\subsection{Cartesian Swap Functions }
By solving for y in equation (1) for  the lower half of the CCMM we get:
\begin{equation}
y=-\sqrt{2lx-x^{2}}+l
\end{equation}
Adding $\Delta_{x}$ and $\Delta_{y}$ into our swap function to get the quantity of $y$ for quantity $x$:
\begin{equation}
\Delta_{y}=-\sqrt{2l(x-\Delta_{x})-(x-\Delta_{x})^{2}}+l-y
\end{equation}
to solve for $\Delta_{x}$ simply rearrange the variables given the function's symmetric nature.

For n=2 CSEMM the swap function becomes:
\begin{equation}
y=-\alpha_y\left(\left(1-\left|\frac{x}{\alpha_x}-1\right|^{\frac{\ln\left(2\right)}{\ln\left(\frac{\alpha_x}{\alpha_x-1}\right)}}\right)^{\frac{\ln\left(\frac{\alpha_y}{\alpha_y-1}\right)}{\ln\left(2\right)}}-1\right)
\end{equation}
Adding $\Delta_{x}$ and $\Delta_{y}$ into our swap function
\begin{equation}
\Delta_{y}=-\alpha_y\left(\left(1-\left|\frac{x+\Delta_{x}}{\alpha_x}-1\right|^{\eta(\alpha_x)}\right)^{\frac{1}{\eta(\alpha_y)}}-1\right)-y
\end{equation}
For $\Delta_{x}$ due to the asymmetric nature of the superellipse our swap function order for $u(x)$ is reversed:
\begin{equation}
\Delta_{x}=-\alpha_y\left(\left(1-\left|\frac{y+\Delta_{y}}{\alpha_x}-1\right|^{\eta(\alpha_y)}\right)^{\frac{1}{\eta(\alpha_x)}}-1\right)-x
\end{equation}

\subsection{Polar Swap in Rust Arbitrum Stylus }

Rust WASM smart contracts from Arbitrum Stylus can help make Solidity contracts more expressive. For example, math in Solidity can be limited. In contrast, Rust has more math function support with its standard library, and has a large Rust cargo collection of math libraries for even more complex math operations. Some of these Rust crate libraries can even be used where the Rust standard library might need to be disabled for compiling WASM binaries.

\subsubsection{Remark 1}
It is recommended to used fixed point math libraries, since floating point calculations can be non deterministic based on the hardware the math is computed on.

\subsubsection{Remark 2}
The Rust math function below uses 
decimal values that aren't scaled. For EVM production deployments, these values should be scaled up in a similar way to how Solidity library prb-math uses WAD values to represent decimal values since the EVM currently only supports integer values.

Rust math function related to the swap function with Line referring to equations outlined in desmos \cite{ref_url18}:

\begin{minted}[fontsize=\small]{rust}
fn get_delta_y_function() -> f64 {
    
    let constant_l: f64 = get_constant_l(); // Line 1
    let l_scaled: f64 = constant_l * 10000.0;
    println!("l_scaled: {}", l_scaled);

    let radians_45_deg = std::f64::consts::FRAC_PI_4;
    let radians_135_deg: f64 = 3.0 * radians_45_deg;
    let cos_expression = radians_135_deg.cos();
    let l_cos_expression = l_scaled * cos_expression;

    let x_in: f64 = 1.0; // Line 6
    let radians_45_deg = std::f64::consts::FRAC_PI_4;
    let radians_135_deg: f64 = 3.0 * radians_45_deg;
    let sin_expression = radians_135_deg.sin();
    let l_sin_expression = l_scaled * sin_expression;
    let numerator = l_sin_expression - x_in;
    let ratio: f64 = numerator / l_scaled;
    let ratio_squared = f64::powf(ratio, 2.0);
    let radicand: f64 = 1.0 - ratio_squared;
    let square_root = radicand.sqrt();
    let l_square_root = l_scaled * square_root;
    
    let output: f64 = l_square_root + l_cos_expression;

    return output; // 0.999958580363; // Line 7
}
\end{minted}

\section{Further Research}
Polar coordinates allow us to add a sinusoidal wave onto the circular invariant. This gives us the ability to construct multimodal liquidity fingerprints with the following equation where $\alpha=[4,...\infty)$ in intervals of $2$ and $\beta=\alpha^2$
\begin{equation}
r\left(\theta\right)=\frac{L}{\sqrt[\beta]{1-\frac{1}{2}\sin\left(\alpha\theta\right)^{2}}}.
\end{equation}

At $\alpha=4$ our trading function resembles Curve \cite{ref_url2}. At $\alpha=6$ it resembles a bimodal liquidity fingerprint the middle part of which can be compared to convex liquidity - liquidity that grows as we move away from the current price \cite{ref_url19}. We would like to particularly draw attention to $\alpha=8$ where we have a trimodal liquidity fingerprint which can be very useful for capturing dynamics of collateralized debt position (CDP) stablecoin such as DAI. We can see empirically that such stablecoins exhibit sinusoidal peaks at $0$, $1\%$, and $-1\%$ when minting and burning are activated [Figure~\ref{fig7}]. We provide an interactive model for further research for such CDP stablecoins in desmos \cite{ref_url20}.

\section{Disclaimer}
This paper is for general information purposes only. It does not constitute investment advice or a recommendation or solicitation to buy or sell any investment and should not be used in the evaluation of the merits of making any investment decision. It should not be relied upon for accounting, legal or tax advice or investment recommendations.

\begin{figure*}[ht!]
\centerline{\includegraphics[width=0.73\textwidth, keepaspectratio]{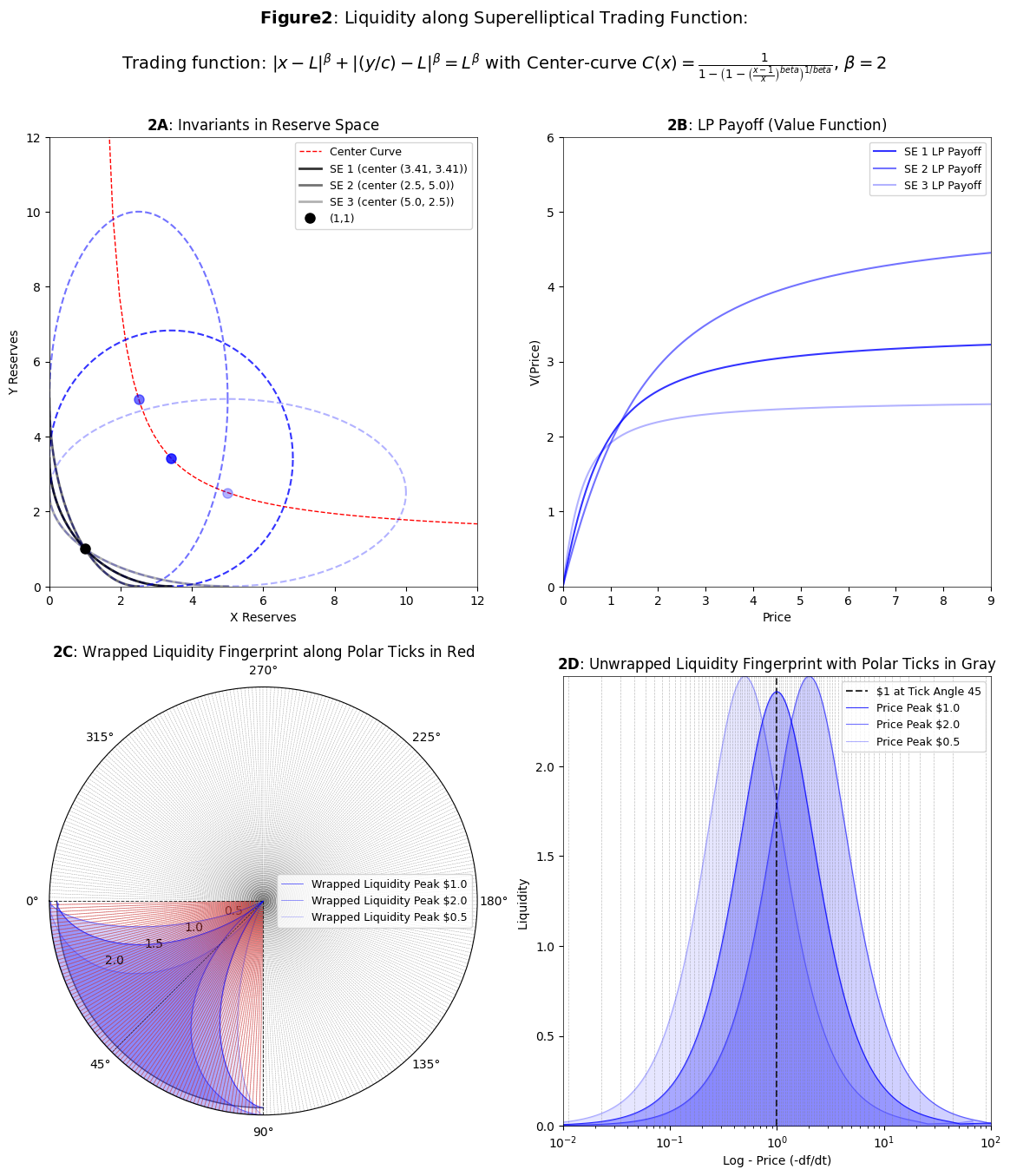}}
\caption{ 2A: The Center-curve in red allows us to shift the center of our superellipse while retaining the trading function's scale invariance as demonstrated by having them cross through coordinate (1,1). 2B: The LP Payoff is given to us via the Legendre Transform. 2C: We can think of liquidity living in polar tick space where each angle represents a particular level of liquidity. 2D:  If we were to unwrap the liquidity, we would observe that most of the polar ticks are concentrated at the center of the body of a symmetric liquidity fingerprint, making such a polar discretization a very good building block for liquidity concentration with ticks.}
\label{fig2}
\end{figure*}

\begin{figure*}[ht!]
\centerline{\includegraphics[width=0.73\textwidth, keepaspectratio]{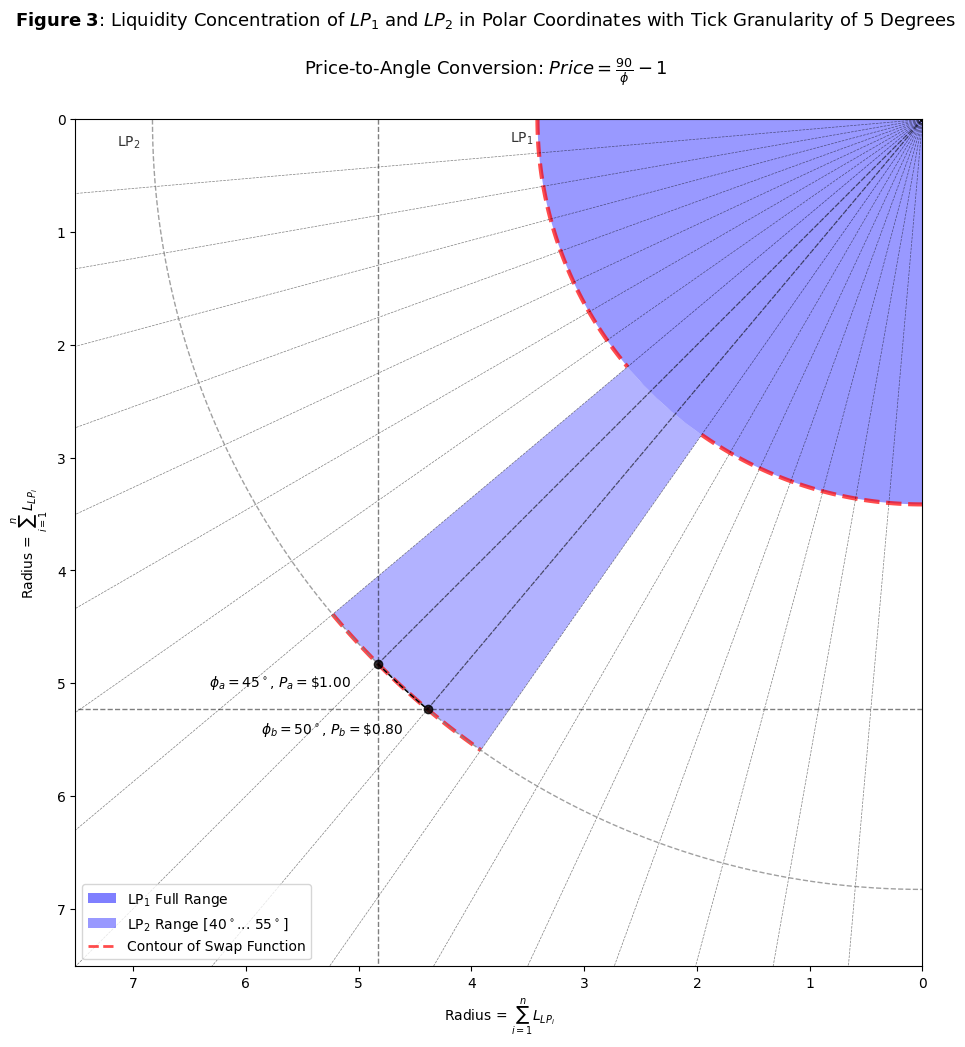}}
\caption{ $LP_1$ provides liquidity in the polar range of $[0...90]$, corresponding to a price range of $[0...\infty)$, while $LP_2$ provide liquidity between the angle range $[40...55]$. The swap function from the price  $P_a$ of \$1 to $P_b$ of \$0.80 can move via  polar coordinates. The red contour of the liquidity distribution is where the swap function resides, which expands and contracts based on the amount of liquidity different LPs have concentrated along each polar tick range. To convert from tick angle to price the formula is reverted: $\frac{90}{\phi}-1=P$.}
\label{fig3}
\end{figure*}

\begin{figure*}[ht!]
\centerline{\includegraphics[width=0.73\textwidth, keepaspectratio]{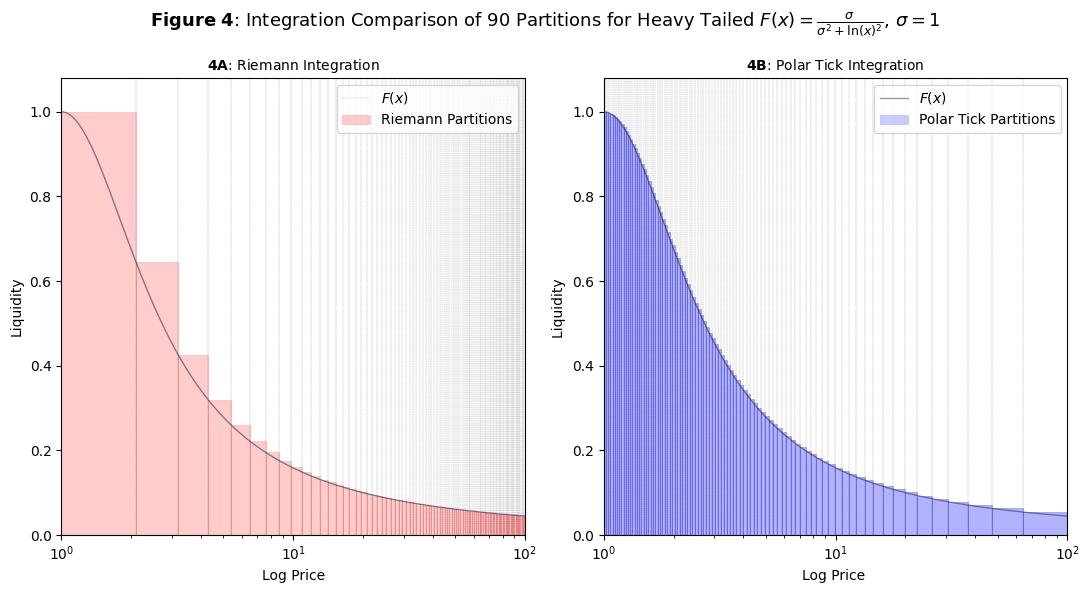}}
\caption{ 4A: Inaccurate capture at the body of the heavy tailed distribution as well as the need for increasing Riemann partitions required to capture the tail with each LP position costing more in gas. 4B: Integration of partitions along polar ticks results in both the body being accurately captured while also capturing the heavy tail.}
\label{fig4}
\end{figure*}

\begin{figure*}[ht!]
\centerline{\includegraphics[width=0.73\textwidth, keepaspectratio]{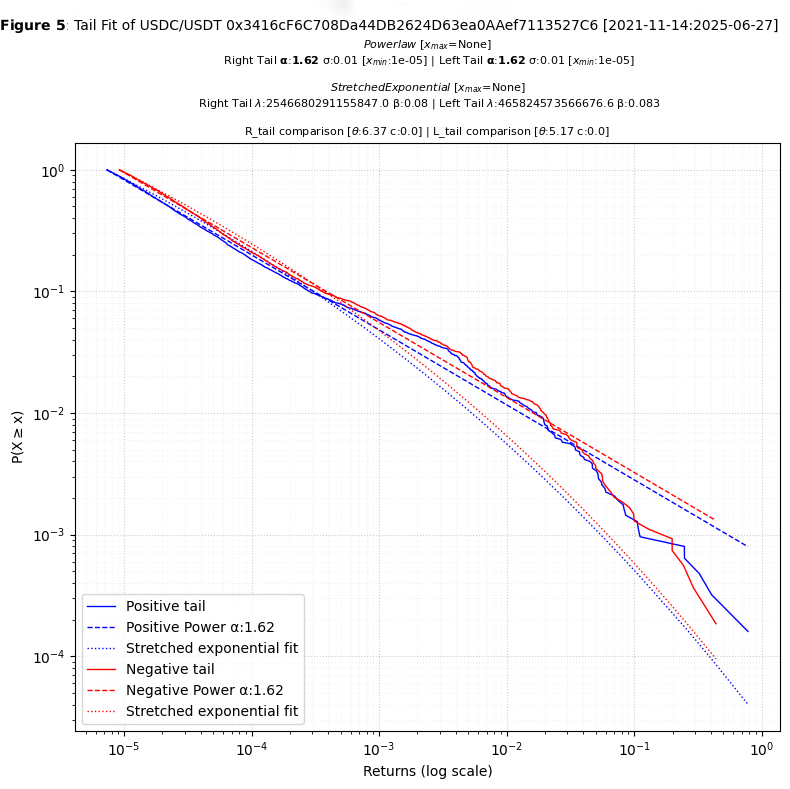}}
\caption{ We have examined the top three most populated and oldest mainnet stablecoin pools on Uniswap v3 retrieved from \href{https://www.dune.com}{Dune Analytics} since inception in rank-frequency plots and found the same pattern for all of them. We have a fat-tail with the power law statistically preferred over the stretched exponential. Its alpha exceeds that of the Cauchy distribution ($\alpha<2$), making the act of stacking multiple such assets into one pool grow in fragility.}
\label{fig5}
\end{figure*}

\begin{figure*}[ht!]
\centerline{\includegraphics[width=0.73\textwidth, keepaspectratio]{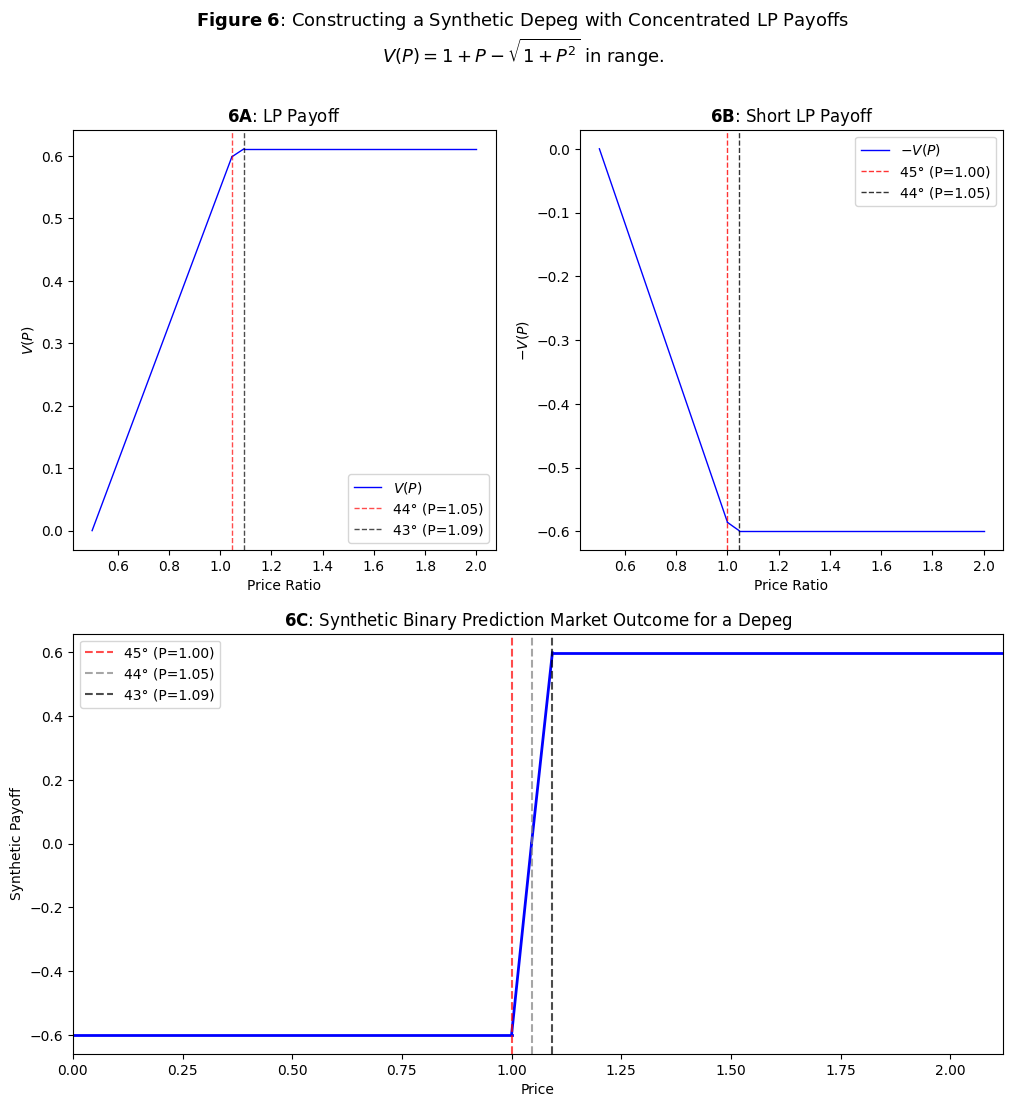}}
\caption{ 6A: Constructing an LP position along a particular degree out of range give us a concave payoff resembling a Uniswap v3 position. 6B: By borrowing, unwrapping, and shorting an LP position below our long position we can have a convex payoff. 6C: combining both positions gives us a binary-like option payoff resembling prediction markets.}
\label{fig6}
\end{figure*}

\begin{figure*}[ht!]
\centerline{\includegraphics[width=0.73\textwidth, keepaspectratio]{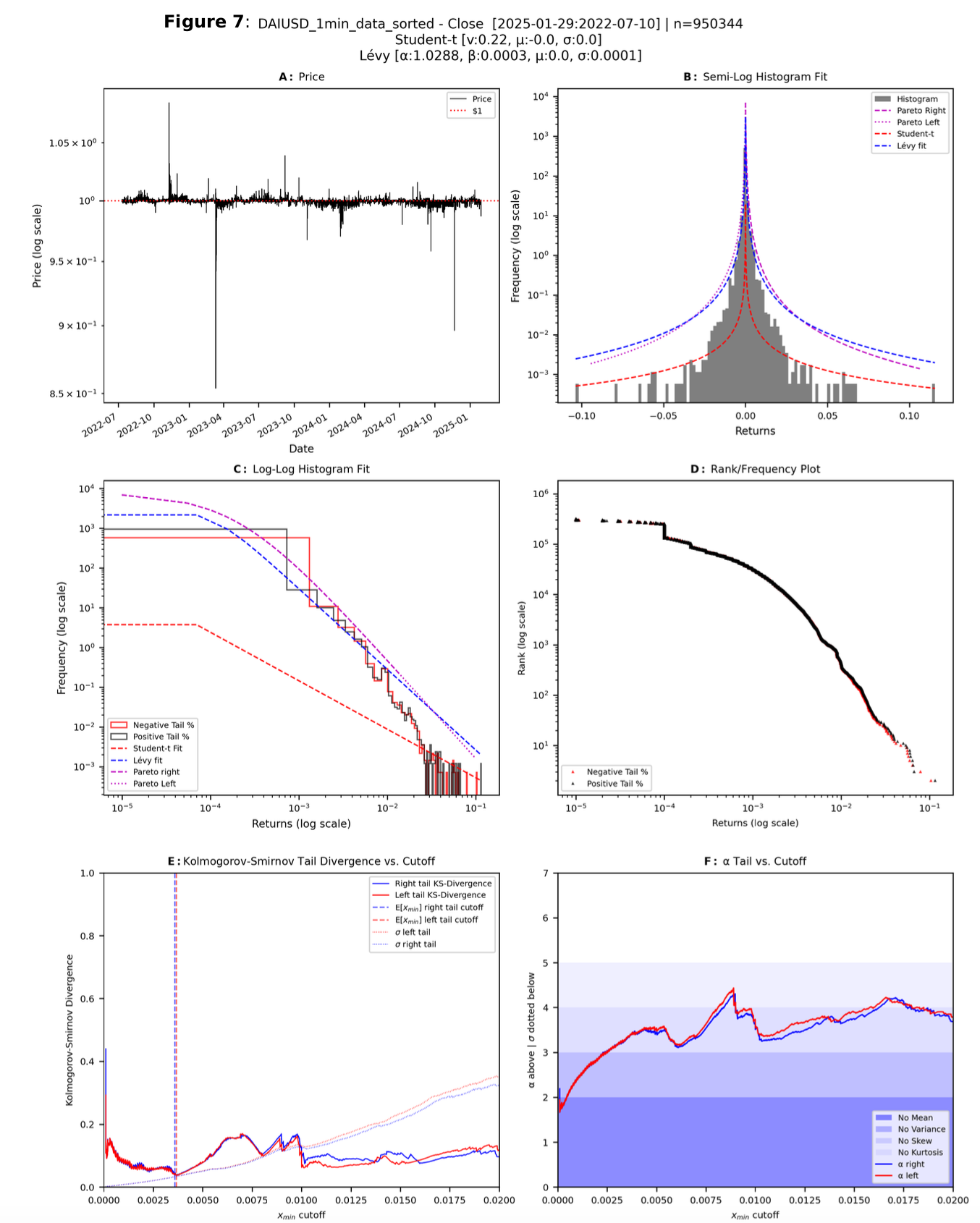}}
\caption{7A: DAI-USD Log-Y price dynamics on a Centralized Exchange (CEX). 7B: Log-Y histogram and fits of
Levy, Student-T and Pareto. For stablecoins, the Pareto distribution provides the
better fit while the Student-T, focusing on outliers, fails to capture the body
giving an inaccurate read. 7C: Log-Log histogram of the tails.
Note the histogram bump at $±1\%$ has a unique sinusoidal bump. 7D: The rank-frequency plot allows us to look more closely at the outliers. There is a unique step function
at the low returns range and a sinusoidal ripple in the tail starting at $1\%$. 7E: We select the earliest minimum $x_{min}$ cutoff value of the Kolmogorov-Smirnov Divergence visualized as dashed verticals in order to minimize our error rate $\sigma$. 7F: We map the power law spectrum showing the fatness of the tail and see the sinusoidal bump prior to $x_{min}$ = $±1\%$.  Data retrieved from \href{https://www.polygon.io}{polygon.io} for CEX data for the price of DAI from the period 2022/07/10 until 2025/01/29 at a one minute time interval and processed in Python.}
\label{fig7}
\end{figure*}

{\ }
{\ }


\begin{thebibliography}{8}
\bibitem{ref_url1}\label{ref_url1}
 Robinson, D: Uniswap v3 - The Universal AMM, \url{https://www.paradigm.xyz/2021/06/uniswap-v3-the-universal-amm}, last accessed 2025/09/19

\bibitem{ref_url2}
 Egorov, M: StableSwap - efficient mechanism for Stablecoin
liquidity, \url{https://docs.curve.finance/assets/pdf/whitepaper_stableswap.pdf}, last accessed 2025/09/19

\bibitem{ref_url3}
 Forgy, E., Lau, L.: A Family of Multi-Asset Automated Market Makers, \url{https://arxiv.org/abs/2111.08115}, last accessed 2025/09/19
 
\bibitem{ref_url4}
 Euler Labs: EulerSwap White Paper, \url{https://github.com/euler-xyz/euler-swap/blob/master/docs/whitepaper/EulerSwap_White_Paper.pdf}, last accessed 2025/09/19

\bibitem{ref_url5}
 Adams, H et al: Uniswap v3 Whitepaper, \url{ https://uniswap.org/whitepaper-v3.pdf}, last accessed 2025/09/19

\bibitem{ref_url6}
 White, D., Robinson, D., Moallemi, C.: Orbital, \url{ https://www.paradigm.xyz/2025/06/orbital}, last accessed 2025/09/19

\bibitem{ref_url7}
 v-for-vasya: Negative Price AMM, \url{https://ethresear.ch/t/negative-price-amm/17833}, last accessed 2025/09/19

\bibitem{ref_url8}
 Tolstikov, V.: Concentrated Superelliptical Market Maker, \url{https://arxiv.org/abs/2410.13265}, last accessed 2025/09/19

\bibitem{ref_url9}
 Schiarizzi, J. Tolstikov, V.: Orbswap, \url{https://github.com/cupOJoseph/orbswap}, last accessed 2025/09/19

\bibitem{ref_url10}
 Wang, Y.: Automated Market Makers for Decentralized Finance (DeFi), \url{https://arxiv.org/abs/2009.01676}, last accessed 2025/09/19

\bibitem{ref_url11}
  Tolstikov, V.: Desmos - Elliptical Orbswap VT, \url{https://www.desmos.com/calculator/ndckdhzbc5}, last accessed 2025/09/19

\bibitem{ref_url12}
 Angeris, G., Evans, A., Chitra, T.: Replicating Market Makers, \url{https://arxiv.org/pdf/2103.14769}, last accessed 2025/09/19

\bibitem{ref_url13}
  Tolstikov, V.: Desmos - Orbswap beta2 c1 VT, \url{https://www.desmos.com/calculator/pmeynkqexn}, last accessed 2025/09/19

\bibitem{ref_url14}
  University of Washington School of Law: The Silicon Valley Bank Collapse \url{https://www.law.uw.edu/news-events/news/2023/svb-collapse}, last accessed 2025/09/19

\bibitem{ref_url15}
  Taleb, N.: Youtube - MINI LECTURE 14 A First Course on Fragility, Convexity, and Antifragility, \url{https://www.youtube.com/watch?v=ovEPIQR65hc}, last accessed 2025/09/19

\bibitem{ref_url16}
 Lambert, G., Kristensen, J.: Panoptic: the perpetual, oracle-free options protocol, \url{https://arxiv.org/abs/2204.14232}, last accessed 2025/09/19

\bibitem{ref_url17}
  Tolstikov, V.: Desmos - CCMM and CSEMM, \url{https://www.desmos.com/calculator/wy74cucie0}, last accessed 2025/09/19

\bibitem{ref_url18}
  Tolstikov, V.: Desmos - Swap in Polar Coordinates VT, \url{https://www.desmos.com/calculator/xxeeuzbvvp}, last accessed 2025/09/19

\bibitem{ref_url19}
  Wentz, M., Tolstikov, V.: Curvy - Custom Curve, \url{https://ethglobal.com/showcase/curvy-vjk7e}, last accessed 2025/09/19

\bibitem{ref_url20}
  Tolstikov, V.: Desmos - Polarswap with Multimodal Liquidity VT, \url{https://www.desmos.com/calculator/vrsexjue5o}, last accessed 2025/09/19

\end{thebibliography}
\end{document}